\newcommand{\dgree}   {\mbox{$ ^\circ                                      $}}

\documentclass[]{raa}

\usepackage{graphicx,times}             
\usepackage{natbib} 
\usepackage{lscape}
\usepackage{rotating}
\usepackage{subfig}
\usepackage{placeins}
\usepackage{hyperref}
\usepackage{txfonts}

\begin{document}

   \title{Updating the orbital ephemeris of the dipping source \\XB 1254-690 and the distance to the source}

   \volnopage{Vol.0 (200x) No.0, 000--000}      
   \setcounter{page}{1}          

   \author{A. F. Gambino
      \inst{1}
   \and R. Iaria
      \inst{1}
   \and T. Di Salvo
      \inst{1}
   \and M. Matranga
      \inst{1}
   \and L. Burderi
      \inst{2}
   \and F. Pintore
      \inst{3}     
  \and A. Riggio
      \inst{2}
   \and A. Sanna
      \inst{2}
   }

   \institute{Universit\`a degli Studi di
  Palermo, Dipartimento di Fisica e Chimica, via Archirafi 36 - 90123 Palermo, Italy; {e-mail: \it angelofrancesco.gambino@unipa.it}\\
        \and
             Universit\`a degli Studi di Cagliari, Dipartimento di Fisica, SP
           Monserrato-Sestu, KM 0.7, 09042 Monserrato, Italy  \\
        \and
              INAF-Istituto di Astrofisica Spaziale e Fisica Cosmica - Milano, via E. Bassini 15, I-20133 Milano, Italy\\
   }

   \date{Received~~2009 month day; accepted~~2009~~month day}

\abstract{ XB 1254-690 is a dipping low mass X-ray binary system hosting a neutron star and showing type I X-ray bursts. 
We aim at obtaining more accurate orbital ephemeris and at constraining the orbital period derivative of the system for the first time. In addition, we want to better constrain the distance to the source in order to locate the system in a well defined evolutive scenario.
We apply for the first time  an orbital timing technique to XB 1254-690, using the arrival times of the dips present in the light curves that have been collected during 26 years of X-ray pointed observations performed from different space missions.
We estimate the dip arrival times using a statistical method that weights the count-rate inside the dip with respect to the level of the persistent emission outside the dip. We fit the obtained delays as a function of the orbital cycles both with a linear and a quadratic function. 
We infer the orbital ephemeris of XB 1254-690 improving the accuracy of the orbital period with respect to previous estimates.
We infer a mass of M$_{2}=0.42\pm 0.04$ M$_{\odot}$ for the donor star, in agreement with the estimations already present in literature, assuming that the star is in thermal equilibrium while it transfers part of its mass via the inner Lagrangian point, and assuming a neutron star mass of 1.4 M$_{\odot}$. Using these assumptions, we also constrain the distance to the source, finding a value of 7.6$\pm 0.8$ kpc. Finally, we discuss the evolution of the system suggesting that it is compatible with a conservative mass transfer driven by magnetic braking.
}

   \authorrunning{A. F. Gambino, et al. }            
   \titlerunning{XB1254-690: reviewing the ephemerides and distance }  

   \maketitle

%
%
\section{Introduction}           
\label{sect:intro}

XB 1254-690 is a persistent low mass X-ray binary (LMXB) showing type I X-ray bursts and dips.
The source coordinates have been accurately derived from \cite{Iaria_chandra} who, using one \textit{Chandra} observation, located the source at RA (J2000)=194.4048\dgree and DEC (J2000)=-69.2886\dgree with a 90\% confidence level error box of 0.6" radius.
The dipping activity has been revealed during an \textit{EXOSAT} observation in 1984 \citep[][]{Courvoisier} and consists in a periodic decrease of the count-rate. This decrease is caused by photoelectric absorption of part of the X-ray emission by the cold (and/or partially ionised) bulge of matter that forms as a consequence of the impact of the transferred plasma from the companion star onto the outer accretion disk \citep[][]{White}.
While the type I X-ray bursts testify the presence of a neutron star (NS) in the system \citep[][]{Mason}, the presence of dips together with the absence of eclipses in the light curve constrain the inclination angle $i$ of the system with respect to the line of sight to the observer between 60\dgree and 80\dgree \citep[see][]{Frank}.
The optical counterpart has been identified by \cite{Griffiths} with the faint V$\simeq$19 star GR Mus.\\   
Timing analysis of the periodic dips led to the estimation of the orbital period of the system. 
\cite{Courvoisier} estimated the orbital period of the system on the basis of the recurrence time of the X-ray dips obtaining a period of 0.162(6) days. \cite{Motch} improved this estimate giving a period of 0.163890(9) days obtained from optical data, while \cite{Diaz} obtained an orbital period of 0.16388875(17) days, performing a periodogram on ASM data.
In their survey of the timing properties of several LMXB systems, instead, \cite{Levine} assigned an orbital period of 0.1638890(4) days to XB 1254-690.
Observations, however, revealed that the dipping activity is not always present and that the dips are quite different in shape from one observation to another. \\
\cite{Battacharyya} reported a weak evidence of quasi periodic oscillations at about 95 Hz detected during one thermonuclear X-ray burst.
\cite{Diaz}, on the basis of the results of their spectral analysis performed on the source using \textit{XMM-Newton} and \textit{INTEGRAL} data, proposed the presence of a tilted accretion disk in the system. In this way they justified the disk temperature changes observed between the dip and non-dip time intervals of the XMM/INTEGRAL light curves, as well as the optical modulation observed with data collected from the optical Monitor (OM) of \textit{XMM-Newton}.
The modulation of the optical light curve was already observed by \cite{Motch}. These authors observed that the optical light curve shows minima occurring 0.15 in phase after the X-ray dips, suggesting that the modulation is due to the varying aspect of the X-ray heated atmosphere of the donor star. They, however, did not rule out the presence of an asymmetric accretion disk that does not completely shadow the companion star.
A further confirmation of the hypothesis of a tilted accretion disk has been proposed by \cite{Cornelisse}. They actually took advantage from the ephemeris of \cite{Diaz}, revealing the presence of a negative superhump (i.e. a periodic photometric hump having a period shorter than the orbital period by a few percent). This supports the idea that XB 1254-690 could host a precessing accretion disk with a retrograde precession motion with a period of (6.74$\pm$0.07) days. \\
\cite{Cornelisse} proposed that the disk is tilted out of the orbital plane along its line of nodes implying that a large fraction of matter transferred from the companion star to the neutron star overflows or underflows the accretion disk instead of hitting onto the disk rim. In their opinion this could explain the presence of the absorption features observed by \cite{Iaria_chandra} and \cite{Boirin}. 
They also found a marginal evidence of a possible positive superhump suggesting that the accretion disk is possibly eccentric due to effects of tidal resonance. On the basis of these results they inferred the mass ratio of the system, q=M$_{2}$/M$_{1}$=0.33-0.36, and constrained the mass of the neutron star, M$_{1}$, between 1.2 and 1.8 M$_{\odot}$.\\

A first estimate of the distance was advanced by \cite{Courvoisier}. Using \textit{EXOSAT} data, they inferred the distance to the source from two type-I bursts, assuming that the luminosity at the burst peak was the Eddington luminosity for a 1.4 M$_{\odot}$ neutron star. They obtained a distance of 12$\pm$2 kpc and 11$\pm$2 kpc for the two bursts, respectively.
Subsequently, \cite{Motch} constrained the distance to the source in the range 8-15 kpc from a modeling of the optical emission. Their modeling also showed that the optical brightness of the source is well explained when assuming that the donor star is near the main sequence.
\cite{Zand} reported an estimation of the distance to the source analysing data collected from \textit{BeppoSAX} in 1999. The detection of a possible photospheric radius expansion (PRE) during one superburst precursor allowed them to estimate a distance of 13$\pm$3 kpc. 
A successive work by \cite{Galloway} allowed to infer an estimation of the distance, again on the basis of the properties of type I X-ray bursts from the source.    
Analysing data collected from the proportional counter array (PCA) on board the \textit{RXTE} mission, \cite{Galloway} found only a marginal evidence of PRE during the observed type I X-ray bursts, and estimated a distance to the source of 15.5$\pm$1.9 kpc, in the case of a companion star with cosmic abundances, and of 20$\pm$2 kpc, in the case of a pure helium donor star, respectively.

In this work, we update the orbital ephemeris of XB 1254-690 using pointed observations collected by different space missions during a total time span of about 26 years. We constrain the orbital period derivative for the first time, and we give a revised estimate of the distance to the source. We also discuss the mass transfer in the system, suggesting that the system experiences a conservative mass transfer driven by magnetic braking of the companion star.
The paper is structured as follows: in Sect. \ref{sec:observation} we describe the data selection and reduction, in Sect. \ref{sec:analysis} we present the data analysis and the results, and in Sect. \ref{sec:discussion} we discuss the results.

\section{Observation and data reduction}
\label{sec:observation}

\begin{table*}
\caption{Observations used for the timing analysis} \label{tab:obs_log}   
     
\centering          
\scriptsize
\begin{tabular}{lccccc}    
\hline\hline       
Sequential n. &
Satellite/Instrument &
Observation ID &
Start time  & 
Stop time  &
Number \\

 &
 &
 &
(UT) & 
(UT) &
 of dips \\

\hline

1 &
EXOSAT/ME &
18332 &
1984 Feb 5 06:07:57 & 
1984 Feb 5 09:08:14 &
1 \\

2 &
EXOSAT/ME &
31571 & 
1984 Aug 7 03:10:27 &
1984 Aug 7 09:59:21 &
2 \\

3 &
EXOSAT/ME &
31593  & 
1984 Aug 7 10:29:26 &
1984 Aug 7 14:04:14 &
1 \\

4 &
EXOSAT/ME &
49647  & 
1985 Apr 15 04:27:23 &
1985 Apr 15 07:57:51 &
1 \\

5 &
Ginga/LAC &
 900802113648, 900803061648 &
1990 Aug 02 11:37:52 &
1990 Aug 03 10:03:08 &
5 \\

6 &
XMM/Epic-pn &
60740101 &
2001 Jan 22 15:48:48 &
2001 Jan 22 20:02:19 &
1 \\

7 &
RXTE/PCA &
60044-01-01-02 &
2001 May 9 23:26:08  & 
2001 May 10 01:37:04 &
1 \\

8 &
RXTE/PCA &
60044-01-01-03, 60044-01-01-05, 60044-01-01-08 &
2001 May 11 17:30:40  & 
2001 May 12 11:46:40 &
3 \\

9 &
XMM/Epic-pn &
405510401 &
2007 Jan 14 01:12:04  & 
2007 Jan 14 18:19:31  &
4 \\

10 &
RXTE/PCA &
93062-01-01-000 &
2008 Jan 16 05:28:00 & 
2008 Jan 16 13:27:44 & 
1 \\

11 &
RXTE/PCA &
95324-01-01-010 &
2009 Dec 31 03:51:12  & 
2009 Dec 31 1 11:50:56  &
1 \\

12 &
RXTE/PCA &
95324-01-02-000, 95324-01-02-00 &
2010 Jan 1 01:44:48  & 
2010 Jan 1 11:33:04  &
3 \\

13&
RXTE/PCA &
95324-01-01-00 &
2009 Dec 31 21:01:52  & 
2009 Dec 31 23:31:44  &
1 \\

\hline
\hline
              
\end{tabular}

\end{table*}
To perform the data analysis of XB 1254-690 we take advantage of all the available pointed observations in X-ray archival data.
However, no dip could be found in \textit{BeppoSAX}, \textit{ASCA}, \textit{ROSAT}, \textit{Swift}, and \textit{Chandra} observations. We therefore analysed only the data collected by \textit{EXOSAT}, \textit{Ginga}, \textit{RXTE} and \textit{XMM-Newton}, which altogether span a temporal window of about 26 years (from 1984 to 2010).\\
\textit{EXOSAT} observed XB 1254-690 eight times between 5 February 1984 and 15 April 1985. We used the background-subtracted data products of the \textit{EXOSAT} Medium Energy experiment (ME) in the energy range between 1 and 8 keV. We binned the light curves at 6 s. In addition, we performed the barycentric corrections using the tool \verb|earth2sun| and giving as an input the coordinates of the source found by \cite{Iaria_chandra}. These coordinates will be used hereafter for all the subsequent analysis.\\

\textit{XMM-Newton} observed the source five times between 22 January 2001 and 9 March 2007 with the EPIC-pn camera. The selected observations have been performed in fast timing mode. We processed the dataset with the \verb|epproc| tool of the Scientific Analysis System (SAS) v. 14.0.0. We extracted source events only from a box centered at the RAWX coordinate of the maximum of the photons distribution (RAWX=37), and having a width of 15 RAWX. We extracted the light curves using the \verb|evselect| tool, selecting only events with PATTERN$\leq$4 (single and double pixel events) and FLAG=0 to ignore spurious events. The light curves have been extracted between 0.5 and 10 keV with a bin time of 0.006 s. The barycentric corrections have been applied with the \verb|barycen| tool.\\

The available observations performed from RXTE/PCA are sparsely distributed in a temporal window of about 13 years (from 1997 to 2010). We used background-subtracted Standard 2 light curves covering the energy range 2 -- 9 keV. The light curves have a bin time of 16 s and the barycentric corrections have been performed using the tool \verb|faxbary|.\\

\textit{Ginga} observed the binary system with the Large Area Counter experiment (LAC) on 17 July 1989 (ObsID 900802113648) and on 3 August 1990 (ObsID  900803061648). We use the background-subtracted light curves collected from the top layer of the detector covering the 2-17 keV energy band. The light curves have a bin time of 16 s and the barycentric corrections have been applied with the ftool \verb|earth2sun|.\\

\section{Data analysis}
\label{sec:analysis}

To obtain the orbital ephemeris of XB 1254-690 we need the arrival times of the dip as a function of time or of the orbital cycle. 
For the first time we apply a timing technique to the arrival times of the dips in order to improve the ephemeris of the source. The ephemeris obtained so far for XB 1254-690 in the X-ray band are based on periodograms performed on ASM data that have a statistic surely lower than the pointed observations we use in our analysis, which altogether span a temporal period of 26 years. \\
To obtain the dip arrival times we have to take into account the fact that XB 1254-690 shows dips varying in shape from one orbital cycle to another.
For this reason, we cannot fit the dip with a specific function since this implies the assumption that the dip has always the same shape \citep[see][in preparation]{Gambino, Iaria_2017}.\\
To address this issue, we take advantage from the method developed by \cite{Hu} to parameterize the dipping behaviour of XB 1916-053, and to systematically study its variation.
The method can be applied both to dippers and eclipsing sources, and represents a powerful tool to obtain the dip arrival time for sources showing dips strongly variable in width and depth during different orbital cycles. The only constraint of this method is that the observation of at least a complete dip is required \citep{Hu}.\\
Therefore we selected all the available pointed observations in which single dips appear to be complete, collecting a total of 14 dips to analyse. In addition to these dips, the Ginga observations (ObsIDs 900802113648 and 900803061648) show 5 incomplete dips. Similarly, three RXTE observations (ObsIDs 60044-01-01-03, 60044-01-01-05 and 60044-01-01-08) show a total of three partial dips close in time, and other two incomplete dips are visible in two RXTE observations (ObsIDs 95324-01-02-000 and 95324-01-02-00). We will take into account all these incomplete dips in the second part of the data analysis.  We report all the selected observations in Table \ref{tab:obs_log}. 
In each light curve we excluded all the type I X-ray bursts present, removing temporal intervals starting 5 s before the rise time and ending 100 s after the peak time of each burst. 
We implemented the method of \cite{Hu} on the available complete dips to find the dip arrival times. For each of these light curves we distinguished between the dip and the persistent (non-dip) states, by roughly guessing the boundaries of the dip (see Table \ref{tab:boundary}). Then, we identified the persistent count-rate of the source by fitting the data points belonging to the persistent state with the linear function that minimises the $\chi^{2}$. Hereafter, the persistent count-rate level will be denoted by I$_{0}$. To obtain the dip arrival time in the light curve of each of the complete dips we have to average the times of each point in the dip 
weighting them by the difference between the corresponding count-rate and that of the predicted persistent state I$_{0}$. 
Then, re-arranging the relation of \cite{Hu} the time  elapsed from the beginning of the  observation at which the dip occurs is given by:
\begin{equation}\label{eq:ph_dip}
t_{dip}=\frac{\sum\limits_{i=1}^N(I_{0}-I_{i})\;t_{i}}{\sum\limits_{i=1}^N(I_{0}-I_{i})}
\end{equation}
where $i$ is an integer index running from the left boundary ($i=1$) to the right boundary ($i=N$) of the dip and I$_{i}$ is the value of count-rate at each time $t_{i}$ included in the dip state domain. From eq. \ref{eq:ph_dip} it is evident that the choice of the dip state boundaries (i=1 and i=N) can be arbitrary. This is widely demonstrated by \cite{Hu} with different tests, and could be naively explained with the fact that the points lying in the persistent state beside the dip give little contribution to the sum (i.e.\ $I_{0}-I_{i} \simeq 0$), and hence to the determination of the time $t_{dip}$ at which the dip occurs. \\
The arrival times of the dips are then calculated as $T_{dip }=T_{start}+t_{dip}$, where $T_{start}$ is the starting time of the specific observation (see Table \ref{tab:obs_log}). We evaluate the delays of the observed dip arrival times with respect to the arrival times predicted using the orbital period $P=14160.01 s$ of \cite{Levine} and an arbitrary reference epoch $T_{0}=12733.0546 \;TJD$, corresponding to the arrival time of the dip observed in the XMM-Newton observation (ObsID 60740101). The arrival times, as well as the corresponding orbital cycle and the delays (O-C) are reported in Table \ref{tab:dip_times} (First Iteration).

\begin{table}
\caption{Determination of the arrival times of complete dips and phases of folded dips.} \label{tab:boundary}   
     
\centering          
\scriptsize
\begin{tabular}{cllcccccc}    
\hline\hline       

Point &
Satellite/Instrument &
ObsID &
Time interval (s) &
Phase Interval &
Dip boundary (s) &
Dip boundary (Phase) &
$t_{dip} \;\; \rm{(s)}$ &
$\phi_{dip}$ \\

\hline

1 &
EXOSAT/ME &
18332 &     
0 -- 13038 &
-- &
5400 -- 6254 &
-- &
5895.3097 &
-- \\

2 &
EXOSAT/ME &
31571 &     
0 -- 12732 &
-- & 
5500 -- 9350 &
-- &
7170.5962 &
-- \\

3 & 
EXOSAT/ME &
31571 &    
15168 -- 29328 &
-- &
18900 -- 22950 &
-- &
21302.0361 &
-- \\

4 & 
EXOSAT/ME &
31593 &    
1044 -- 15204 &
-- &
6600 -- 10650 &
-- &
8423.0805 &
-- \\

5 &    
EXOSAT/ME &
49647 &
3018 -- 17178 &
-- &
8548 -- 10748 &
-- &
9639.4763 &
-- \\

6 &   
Ginga/LAC &
900802113648, &  
-- &
0.55 -- 1.55 &
-- &
0.91 -- 1.13 &
-- &
1.04 \\

 &   
 &
900803061648 &  
 &
 &
 &
 &
 &
 \\

7 & 
XMM/Epic-pn &
60740101 &    
5542 -- 19702 &
-- &
11301 -- 13701 &
-- &
12259.9043 &
-- \\

8 &   
RXTE/PCA &
60044-01-01-02 &  
0 -- 9368 &
-- &
300 -- 3500 &
-- &
1917.2109 &
 --\\

 9 &     
 RXTE/PCA &
60044-01-01-03, &
-- &
0.47 -- 1.47 &
-- &
0.83 -- 1.06 &
-- &
0.95 \\

  &     
  &
60044-01-01-05, &
 &
 &
 &
 &
 &
 \\

 &    
 &
60044-01-01-08 &
 &
 &
 &
 &
 &
 \\

10 &    
XMM/Epic-pn &
405510401 & 
1277 -- 15437 &
-- &
8150 -- 10051 &
-- &
9143.1486 &
-- \\

11 &    
XMM/Epic-pn &
405510401 & 
15462 -- 29622 &
-- &
21800 -- 23100 &
-- &
22526.2440 &
-- \\

12 &   
XMM/Epic-pn &
405510401 &  
30313 -- 44473 &
-- &
36401.0, 38401.0 &
-- &
37425.4854 &
-- \\

13 &   
XMM/Epic-pn &
405510401 &  
45231 -- 59391 &
-- &
50801 -- 52901 &
-- &
52037.2646 &
-- \\

14 &   
RXTE/PCA &
93062-01-01-000 &  
0 -- 10360 &
-- &
2500 -- 4000 &
-- &
3235.2159 &
-- \\

15 &     
RXTE/PCA &
95324-01-01-010 &
19352 -- 33512 &
-- &
26205 -- 26705 &
-- &
26409.6572 &
-- \\

16 &     
RXTE/PCA &
95324-01-02-000, &
-- &
0.54 -- 1.54 &
-- &
0.88, 1.20 &
-- &
1.05 \\

 &     
 &
95324-01-02-00 &
 &
 &
 &
 &
 &
 \\

17 &     
RXTE/PCA &
95324-01-01-00 &
56 -- 14216 &
-- &
6610 -- 8890 &
-- &
6990.6641 &
-- \\

\hline
\hline
              
\end{tabular}
\tablecomments{0.86\textwidth}{t$_{dip}$ is given in seconds from the start time of the corresponding observation. $\phi_{dip}$ is the phase of arrival derived from the folding of close-in-time incomplete dips.}

\end{table}

\begin{figure}
	 \centering
	 
	\includegraphics[angle=0, width=7.2cm]{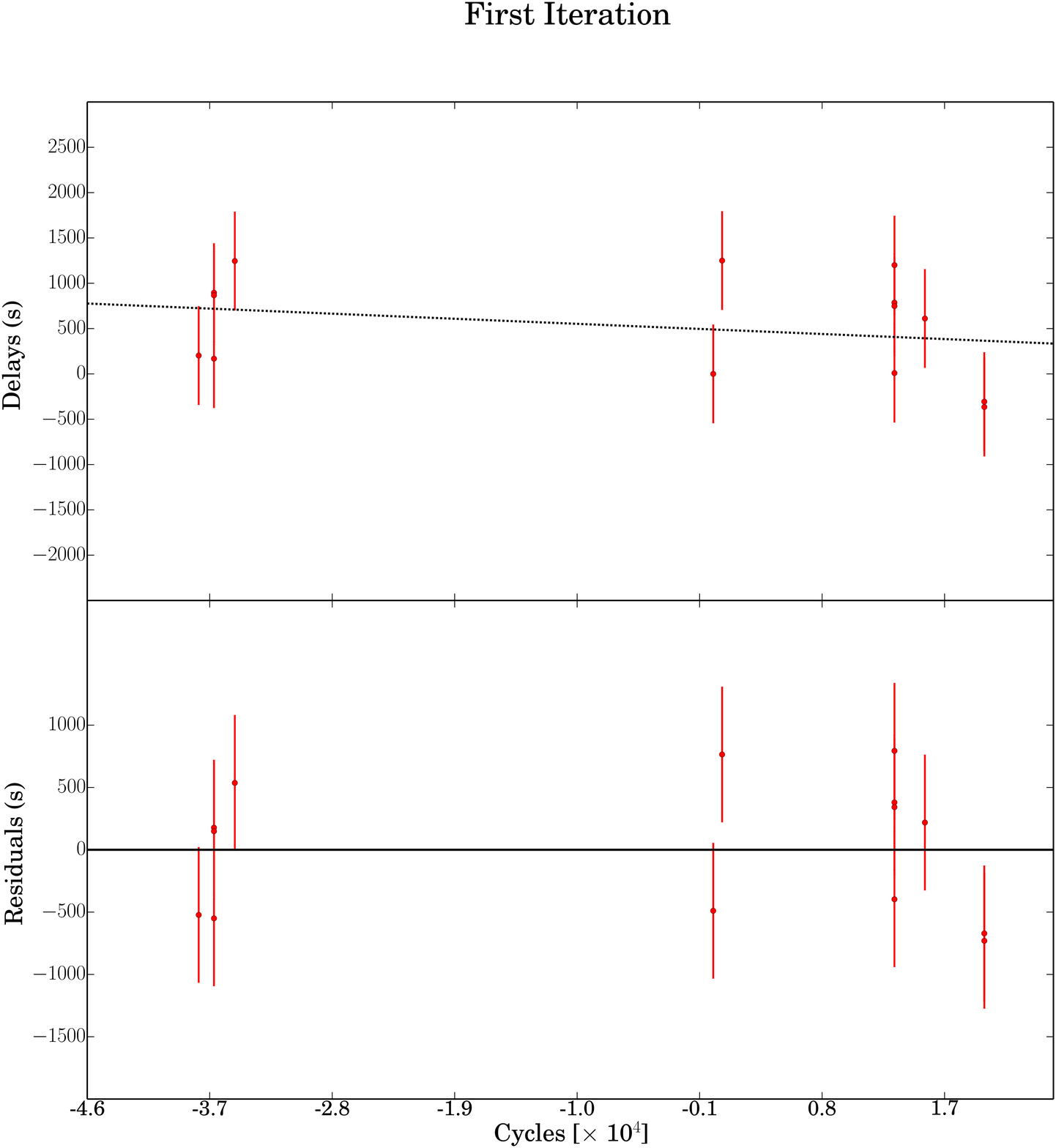}
	 \includegraphics[angle=0, width=7.2cm]{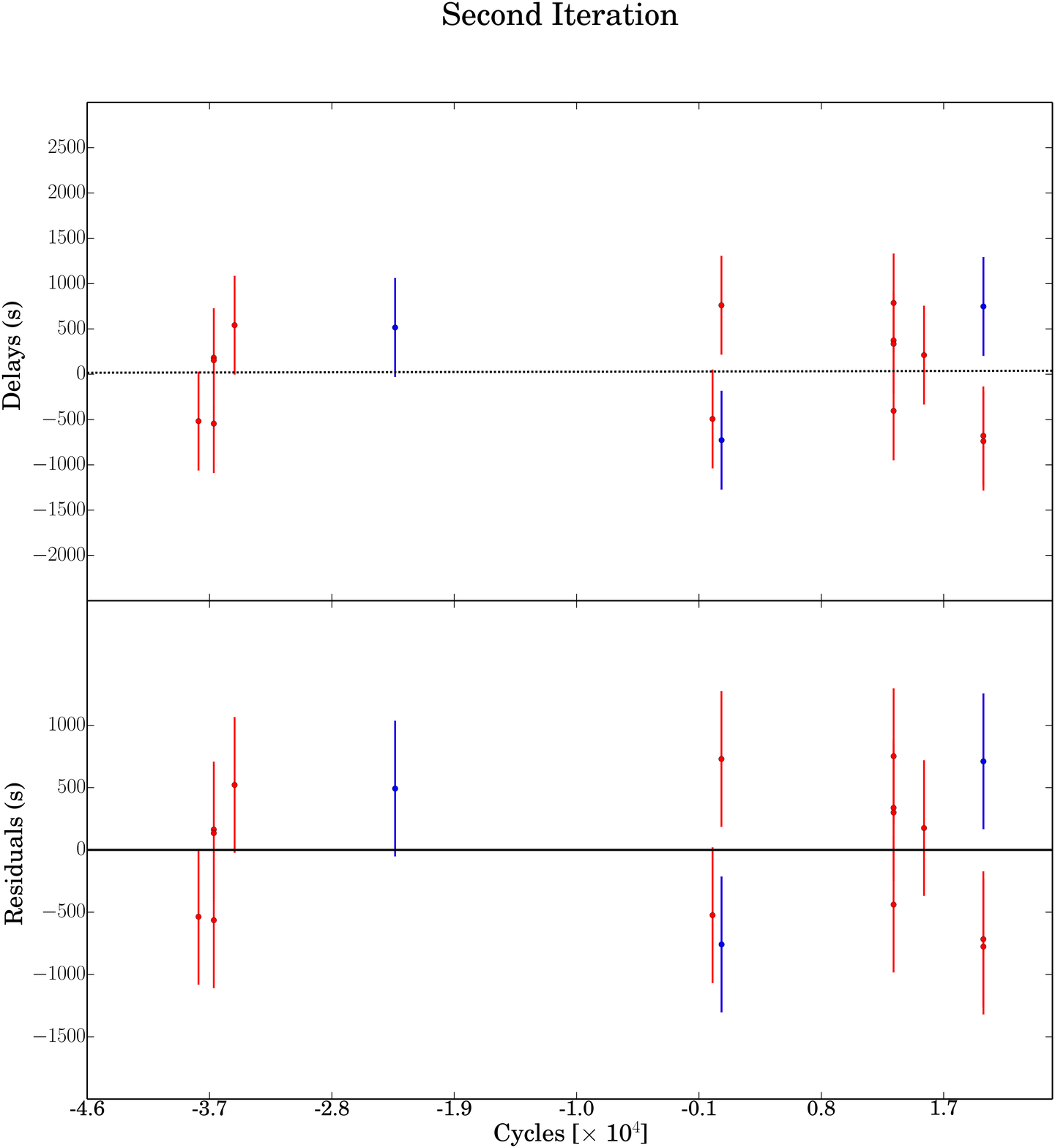}

	 \caption{Delays as a function of the orbital cycle for the first and the second iteration. The blue points represent the supplementary points added in the second iteration. Upper panel: linear fit of the delays.  
	 Lower panel: residuals of the delays with respect to the linear best-fit function.}
	 
	 \label{fig:fit_delays}
	\end{figure}

The error associated with the delays of the dip arrival times is determined by the standard deviation $\sigma$ of the distribution of the obtained phase delays associated to each dip. 
The $\sigma$ of the distribution is equal to 0.04 that corresponds to 544 s according to the trial orbital period we used.\\
Using the same technique performed in \cite{Gambino} and in \cite{Iaria1916}, we fit the delays with a linear function
\begin{equation}\label{eq:f_lin}
y(N)=a+bN,
\end{equation}
where N is the number of orbital cycles, $b$ is the correction to the trial orbital period ($\Delta P_{0}$) in seconds and $a$ is the correction to the trial reference time ($\Delta T_{0}$) in seconds. We obtain $\chi^{2}(d.o.f.)$=  12.94(12). The best-fit model parameters are reported in Table \ref{tab:fit_res} (First Iteration). Applying the obtained corrections for the trial orbital period and the trial reference time we find the following linear orbital ephemeris:
\begin{equation}\label{eq:lin_eph}
T_{dip}(N)= {\rm TJD(TDB)} \; 11931.8065(17) +  \frac{14160.004(6)}{86400} N,
\end{equation}
where 11931.8065(17) TJD and 14160.004(6) s are the new reference time and orbital period, respectively. The associated errors are at 68\% confidence level. \\
However, we expect that, due to the orbital evolution of the binary system, a quadratic term has to be included in the orbital ephemeris. At this purpose, our method easily allows to evaluate the orbital period derivative of the system taking into account the possibility that the delays follow a quadratic trend. Then, we fitted the delays  with a quadratic function as
\begin{equation}\label{eq:f_quad}
y(N)=a+bN+cN^{2},
\end{equation}
where $a$ is the reference time correction ($\Delta T_{0}$) in seconds, $b$ is the orbital period correction ($\Delta P_{0}$) in seconds and $c=\frac{1}{2}P_{0}\dot{P}$ in units of seconds. The best-fit parameters are reported in Table \ref{tab:fit_res} (First Iteration) and allow to obtain a new reference time T$_{0}$ of 11931.812(4) TJD, a new orbital period of 14159.984(15) s and the orbital period derivative $\dot{P}=(-1.3 \pm 2.0)\times 10^{-10} s/s $. Here, the associated error on the orbital period is at 68\% confidence level, while the error on the orbital period derivative is at 95\% confidence level. \\
Nevertheless, this quadratic fit gives a $\chi^{2}(d.o.f.)$=10.73(11) and the estimated F-test probability of chance improvement with respect to the previous linear fit is of the 82\%. This suggests that adopting the quadratic ephemeris does not improve significantly the fit. \\
The delays as a function of the number of orbital cycles are shown in the upper panel of Fig. \ref{fig:fit_delays}. Superimposed we report the best-fit linear function as a solid line. In the lower panel of the same figure we show the residuals of the delays with respect to the linear best-fit function. The maximum deviation of the points with respect to the linear model is of 794 s, that is the 6\% of the orbital period. 
Note, however, that error we associated to the delays of the dip arrival times, while taking into account the statistical error produced by the photon counting and by the phase jitter, also includes the contribution of the linear and quadratic terms, or even of higher orders of the time derivative of the orbital period.
In order to avoid to overestimate the uncertainties on the fitting parameters, we performed again the linear and quadratic fits, assuming the \textit{post-fit} standard deviation as error for each point. This error is determined by the distribution of the points around the best-fit parabolic trend. In this case, therefore, the $\chi^2$ cannot be used as an estimator of the goodness of the fit, because the error is exactly equal to the distribution of the points around the best-fit function, but in this way we get a correct estimate of the uncertainty of the fit parameters.
The fits returned the same parameters we found previously and that are reported in Table \ref{tab:fit_res}. This means that the uncertainty in the fit parameters is dominated in this case by the large scattering intrinsic in the data.\\   
In order to increase the statistics on the timing technique, we can also take the advantage from the observations performed by RXTE and Ginga (sequential numbers 5, 8 and 12 in Table \ref{tab:obs_log}) that we excluded in the first part of the analysis. These pointed observations, in fact, do not show a complete dip in the light curves, but if conveniently folded when close in time, they can provide further measurements of the dip arrival times. \\
We folded each group of these observations using the updated ephemeris given in eq. \ref{eq:lin_eph}.  As already done in the first part of the analysis we distinguished the dip state from the persistent non-dip state guessing the boundaries of the dip. Then, we applied the method of \cite{Hu} to each folded dip profile obtaining the phases at which the dips occur. The dip boundaries as well as the phases at which the folded dips occur are reported in Table \ref{tab:boundary}.\\
The dip arrival times are estimated, starting from the obtained phases, as $T_{dip}=T_{0}+(N+\phi_{dip})P_{0}$, where $T_{0}$ and $P_{0}$ are the reference epoch and the orbital period evaluated with the updated ephemeris of eq. \ref{eq:lin_eph}. With this new trial orbital period $P_{0}$ we obtain the delays of the dip arrival times with respect to $T_{0}$. To be conservative with the first part of the analysis, we associate to these delays the same error that we evaluated for the first set of delays already analysed during the first iteration. In Table \ref{tab:dip_times} (Second Iteration) we report the dip arrival times, the orbital cycles as well as the delays for these supplementary observations.\\
To integrate the delays evaluated in the first part of the analysis with those evaluated just now, we rescaled the delays obtained in the first iteration with respect to the new $T_{0}$ and $P_{0}$ of the updated ephemeris in eq. \ref{eq:lin_eph}. We show the whole set of delays as function of the corresponding orbital cycles in the right panel of Fig. \ref{fig:fit_delays}. We fitted all the delays with respect to the linear function in eq. \ref{eq:f_lin}, obtaining a $\chi^{2}(d.o.f.)$=  17.47(15). We report the best-fit model parameters in Table \ref{tab:fit_res} (Second Iteration). Applying the corrections suggested by the linear fit, we find the following new linear orbital ephemeris:

\begin{equation}\label{eq:lin_eph_bis}
T_{dip}(N)= {\rm TJD(TDB)} \; 11931.8069(16) +  \frac{14160.004(6)}{86400} N,
\end{equation}
where $11931.8069(16) TJD$ and $14160.004(6) s$ are the new corrected reference epoch and orbital period, respectively.
As done before, we also tried to fit the delays with the quadratic function of eq. \ref{eq:f_quad}. 
Also in this case, the best-fit parameters are reported in Table \ref{tab:fit_res} (Second Iteration). Applying the corrections returned by the quadratic fit to the starting ephemeris we obtain a new reference time T$_{0}$ of 11931.808(3) TJD, a new orbital period of 14160.000(11) s and the orbital period derivative $\dot{P}=(0.0 \pm 1.4)\times 10^{-10} s/s$. The associated error on the orbital period is at 68\% confidence level, while the error on the orbital period derivative is at 95\% confidence level. \\
Nevertheless, the quadratic fit gives a $\chi^{2}(d.o.f.)$=17.20(14) and the estimated F-test probability of chance improvement with respect to the previous linear fit is of the 63\%. This suggests that adopting the quadratic ephemeris we do not improve significantly the fit. \\
Again, we ran the same fits using the post-fit standard deviation as error for each delay, obtaining the compatible results. \\
Even though the new ephemeris of eq. \ref{eq:lin_eph_bis} does not significantly improve the ephemeris of eq. \ref{eq:lin_eph}, they are extended on a data set that include all the available pointed observations in the X-ray archive. 
In addition, although the quadratic ephemeris does not improve the significance of the fit, it is fundamental to evaluate an upper limit on the orbital period derivative. The one reported in this paper represents also the only available constraint on the orbital period derivative present to date. This constraint will be improved when further observations of XB 1254-690 will be available.

\begin{figure}
	 \centering
	 
	\includegraphics[angle=0, width=10cm]{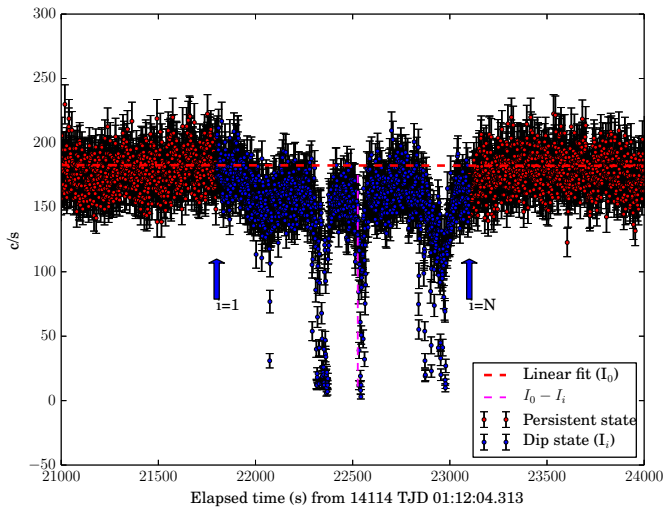}

	 \caption{One of the dips (Point 11) analyses in this work, corresponding to an observation (ObsID 405510401) performed by the \textit{XMM-Newton} space mission.  We indicate the division between the dip state (delimited between i=1 and i=N, respectively) and the persistent state. The linear function fitting the persistent count-rate is shown as a red dashed line (I$_{0}$).}
	 
	 \label{fig:dip}
	\end{figure}

\begin{table}
\caption{Journal of arrival times of the X-ray dips obtained from each light curve for both the iterations performed in the data analysis.   
} \label{tab:dip_times}   
     
\centering          
\scriptsize
\begin{tabular}{cccc|ccc}  
       
& \multicolumn{3}{c}{\textbf{First Iteration}} & \multicolumn{3}{c}{\textbf{Second Iteration}} \\  
\hline

Point &
Dip time &
Cycle &
Delay &
Dip time &
Cycle &
Delay \\

    &
(TJD;TDB) &
    &
(s) &
(TJD;TDB) &
    &
(s) \\

\hline

1 &     
5735.324&
-37809 &
202(544) &
5735.324 &
-37809 &
 -517(544) \\

2 &     
5919.215 &
-36687 & 
896(544) &
5919.215 &
-36687 &
 182(544) \\

3 &     
5919.379 &
-36686 &
867(544) &
5919.379 &
-36686 &
154(544) \\

4 &     
5919.535&
-36685 &
168(544) &
5919.535 &
 -36685 &
-545(544) \\

5&    
6170.297 &
-35155 &
1245(544) &
6170.297 &
-35155 &
541(544) \\

6 &     
-- &
-- &
-- &
11931.762 &
-23347 &
516(544) \\

7 &     
11931.801 &
0 &
0(544)&
11931.801 &
0 &
-493(544) \\

8 &     
12038.999 &
654 &
1250(544) &
12038.999 &
654 &
761(544) \\

9 &     
-- &
-- &
-- &
11931.808 &
665 &
-728(544) \\

10 &     
14114.156 &
13316 &
 786(544) &
14114.156 &
13316 &
373(544) \\

11 &     
14114.311 &
13317 &
 9(544) &
14114.311 &
13317 &
-404(544) \\

12 &     
14114.483 &
13318 &
 748(544) &
14114.483 &
13318 &
335(544) \\

13 &     
14114.652 &
13319 &
 1200(544) &
14114.652 &
13319 &
787(544) \\

14 &     
14481.265 &
15556 &
 611(544) &
14481.265 &
15556 &
211(544) \\

15 &     
15196.466 &
19920 &
 -306(544) &
15196.466 &
19920 &
-680(544) \\

16 &     
-- &
-- &
-- &
11931.844 &
19924 &
748(544) \\

17 &     
15196.957 &
19923 &
 -365(544) &
15196.957 &
19923 &
-739(544) \\
\hline
\hline
              
\end{tabular}

\end{table}

	\begin{table}[h]
\caption{Best-fit values obtained from the linear and quadratic fits of the delays of the dips arrival times.} \label{tab:fit_res} 
     
\centering          
\small

\begin{tabular}{lcc|cc}  
& \multicolumn{2}{c}{\textbf{First Iteration}} & \multicolumn{2}{c}{\textbf{Second Iteration}} \\    
\hline     

Parameter &
Linear &
Quadratic &
Linear &
Quadratic \\

\hline

a (s) &
489 $\pm$ 154 &
954 $\pm$ 342  &
31$\pm$134 &
149$\pm$264\\

b ($\times10^{-3}$ s)&	
-6 $\pm$ 6 &
-26$\pm$15 &
0$\pm$6 &
-5$\pm$11 \\

c ($\times10^{-7}$ s) &	
- &
-9$\pm$6 &
-- &
-3$\pm$5 \\

\hline

\hline
              
\end{tabular}

\end{table}

\section{Discussion}
\label{sec:discussion}

We derived and improved the orbital ephemeris for XB 1254-690 taking the advantage of the whole X-ray data archive, that consists in pointed observations spanning 26 years. The direct measurement of the dip arrival times allowed to increase the accuracy of the orbital period of the system by a factor of 10 with respect to the previous estimation of \cite{Levine}, and by a factor of 4 with respect to the value estimated by \cite{Diaz}. Furthermore, we evaluated for the first time a constraint on the orbital period derivative of $|\dot{P}|<1.4\times 10^{-10}\;s/s$.
This value is compatible with zero and includes both positive and negative values and for this reason should be considered an upper limit on the modulus of the orbital period derivative. However, the result represents a first evaluation of this orbital parameter so far in literature for this source and will be certainly improved including future observations when these will be available.\\
In the following, using the equations for the secular evolution of the source, we discuss the mass transfer for XB 1254-690 in order to get more information on the system. As a first step, we evaluate the observed mass accretion rate onto the neutron star surface.\\
\cite{Iaria_beppo}, modelling the spectrum of XB 1254-690 collected by \textit{BeppoSAX} in the wide band 0.1 - 100 keV, estimated an averaged unabsorbed flux of 1.4$\times 10^{-9}$ erg cm$^{-2}$ s$^{-1}$. 
Taking into account the value of distance of 15.5$\pm$1.9 kpc proposed by \cite{Galloway}, and the value of flux inferred by \cite{Iaria_beppo}, assuming a neutron star radius R$_{NS}$ of 10 km, we can estimate the observed mass accretion rate onto the neutron star as 
\begin{equation}\label{eq:m_dot_oss}
\dot{M}_{obs}=\frac{L_{X}R_{NS}}{G M_{1}}=\frac{4 \pi d^{2} \Phi R_{NS}}{G M_{1}},
\end{equation}
where L$_{X}$ is the observed bolometric X-ray luminosity, M$_{1}$ is the mass of the neutron star that we assumed to be 1.4 M$_{\odot}$ and $\Phi$ is the flux observed by \cite{Iaria_beppo}. We obtain an observational mass accretion rate $\dot{M}$ of (3.4$\pm$0.8)$\times 10^{-9}$ M$_{\odot}$ yr$^{-1}$.\\
To understand in which evolutive scenario the system has to be located, we compare the observed mass accretion rate with that predicted by the theory of secular evolution for low mass X-ray binary systems. As mechanisms of angular momentum loss we take into account the emission of gravitational waves (gravitational radiation), as well as the magnetic braking. 
We do not rule out the possibility that the magnetic braking term plays a role as a mechanism of angular momentum loss, owing to the fact that the mass of the companion star, inferred assuming the condition of thermal equilibrium \citep[see eq. 25 in][]{Verbunt}, is large enough to generate the dynamo effect resulting in a net magnetic field anchored into the companion star surface \citep[see][]{Nelson}.\\
The mass accretion rate predicted by the theory of secular evolution, under the assumed hypothesis, is given by the relation of \cite{Burderi} \citep[see also][]{Di_Salvo2008}:
	 
\begin{figure*}
	 \centering
	 
	 \includegraphics[angle=0, width=7.2cm]{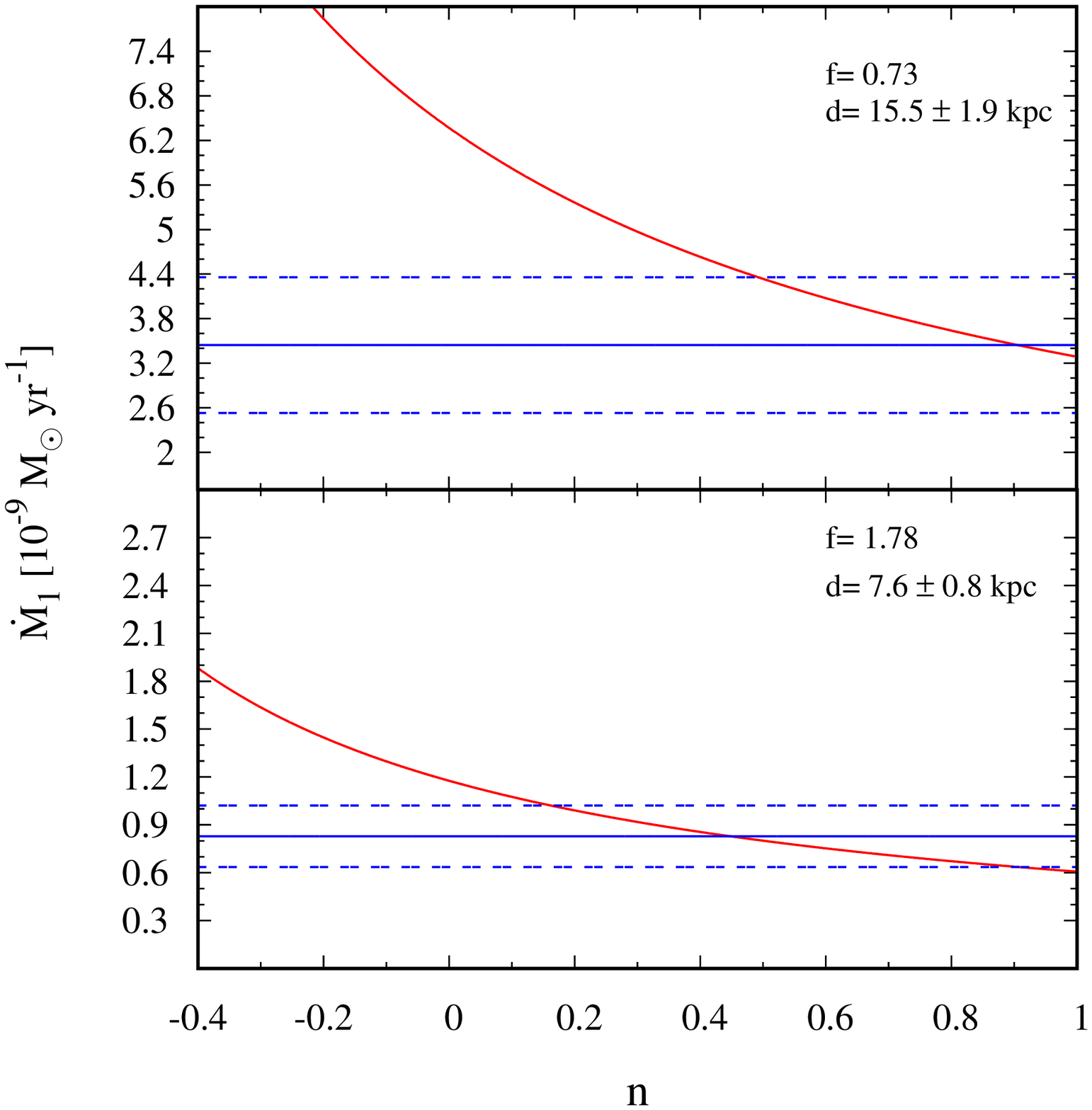}
	 \includegraphics[angle=0, width=7.2cm]{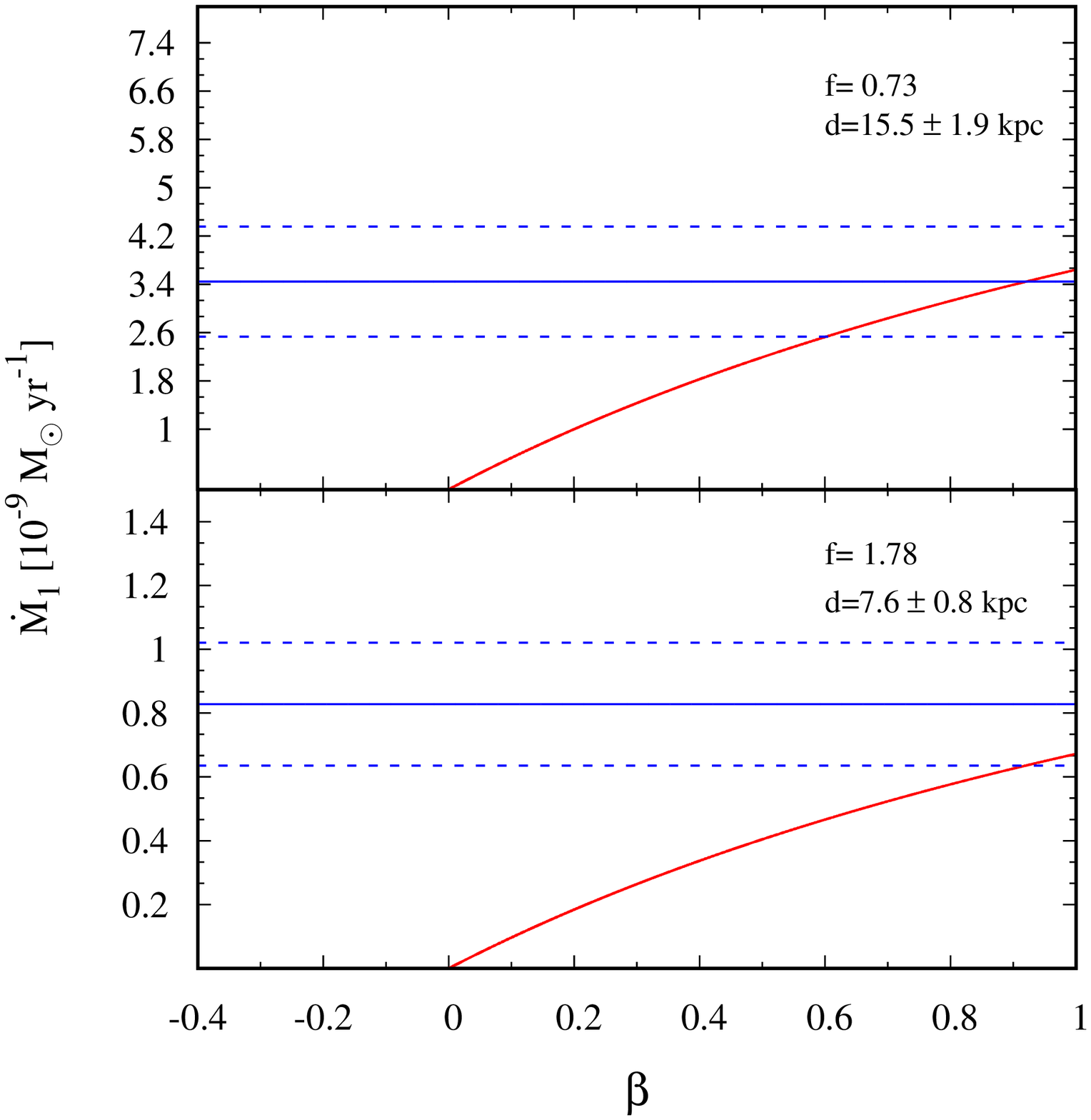}
  
	 \caption{Red curves: theoretical mass accretion rate onto the neutron star surface predicted by eq. \ref{eq:mdotth}. Blue lines: best value of the mass accretion rate obtained from the observations (solid line) and relative errors (dashed lines) for d=15.5$\pm$1.9 kpc and 7.6$\pm$0.8 kpc (upper and lower panels, respectively). \\Left plot: mass accretion rate as a function of $n$ adopting a conservative mass transfer scenario ($\beta$=1). Right plot: mass accretion rate as a function of $\beta$ adopting an index $n$=0.8 (donor star in thermal equilibrium).}
	 
	 \label{fig:confronto_ev}

	 \end{figure*}

\begin{equation}
\label{eq:mdotth}
\dot m_{-8} = - 3.5 \times 10^{-4} [1.0 + {\rm T_{MB}}]
 \; m_1 \; m_{2, \; 0.1}^2 \; m^{-1/3} P_{5h}^{-8/3} \times {\rm F}(n,g(\beta,q,\alpha)),
\end{equation}
where 
$${\rm F}(n,g(\beta,q,\alpha)) = [n - 1/3 + 2g(\beta,q,\alpha)]^{-1}$$
and
$$g(\beta,q,\alpha) = 1 - \beta q - (1-\beta) (\alpha + q/3)/(1+q).$$
In these relations, $\dot{m}_{-8}$ is the secondary mass derivative (negative since the
secondary star looses mass) in units of 10$^{-8}$ M$_{\odot}$ yr$^{-1}$, $m$ is the sum of the masses of the neutron star and of the donor star (m$_{1}$ and m$_{2}$, respectively) in units of solar masses, while m$_{2, \; 0.1}$ is the donor star mass in units of 0.1 solar masses. In addition, $q=m_{2}/m_{1}$, $\beta$ is the fraction of the mass transferred by the companion that is accreted onto the neutron star surface, $\alpha$ is the specific angular momentum of the mass leaving the system in units of the specific angular momentum of the companion star,  P$_{5h}$ is the orbital period in units of five hours and $n$ is the index of the mass-radius relation adopted in eq. \ref{eq:m2_th}. Being associated to the internal structure of the companion star, $n$  can assume values ranging from 0.8 to -1/3. In particular, $n=$0.8 is the index that is associated to stars in thermal equilibrium \citep[see][]{Neece}, meanwhile $n=$-1/3 is associated to stars outside the thermal equilibrium that are also fully convective \citep[see e.g.][]{Burderi}. We suppose, moreover, that the mass expelled from the system (if any) is ejected from the position of the inner Lagrangian point. Then we fix $\alpha=0.7$.\\
The term $\rm T_{MB}$ represents the contribution of angular momentum loss due to the magnetic braking. Re-arranging the expression reported by \cite{Burderi}, this term can be expressed as:
\begin{equation} 
\label{eq:mag_bra}
{\rm T_{MB}} = 8.4 \; k_{0.1}^{2}f^{-2} m_{1}^{-1}q^{1/3}(1+q)^{2/3} P_{2h}^{2},
\end{equation}
where $f$ is a dimensionless parameter of order of unity: preferred values are
$f = 0.73$ \citep{Skumanich} or $f = 1.78$ \citep{Smith}, and $k_{0.1}$
is the radius of gyration of the star in units of $0.1$ \citep{Claret}. Here, revealing in advance one result of the subsequent analysis, we will assume that the mass of the companion star is M$_{2}$=0.42 M$_{\odot}$, which is the mass the companion star, supposed a main sequence star, should have in order to fill its Roche Lobe for the orbital period of the system, $\sim 3.9$ h \citep[see eq. 25 in][]{Verbunt}.\\
We can compare the theoretical mass accretion rate predicted by eq. \ref{eq:mdotth} adopting a conservative mass transfer scenario ($\beta$=1) with the observed mass accretion rate in the source.\\
Adopting a distance to the source of d=15.5$\pm$1.9 kpc \citep{Galloway}, and setting the parameter $f$ of eq. \ref{eq:mag_bra} equal to 0.73 \citep{Skumanich}, we observe an accordance between theory and observations at n$\sim$0.8 (see Fig. \ref{fig:confronto_ev}, left upper panel). This means that according to the theory of secular evolution of the X-ray binary systems, we expect that the companion star is in thermal equilibrium. For the same distance, we also tried the parameter f=1.78 \citep{Smith} for the magnetic braking obtaining that there is accordance neither with n=0.8 nor with n=-1/3.
Furthermore, we explored the case in which the magnetic braking term of eq. \ref{eq:mag_bra} does not play a role in the angular momentum losses of the binary system. In this case, however, we do not observe any accordance between the theoretical mass accretion rate and that observed by \cite{Iaria_beppo}, owing to the fact that in this case the theoretical mass accretion rate underestimates the observed mass accretion rate for any value of $n$. \\
This value of distance, however, as well as the distance of d=13$\pm$3 kpc inferred by \cite{Zand}, has been inferred by the analysis of X-ray type-I bursts, assuming that they show photospheric radius expansion (PRE), and thus, that the peak luminosity of the bursts is the Eddington luminosity for a neutron star of 1.4 M$_{\odot}$. However, as the same authors reported in their works, the observations they analysed have not enough statistics to be sure of a PRE, and as a consequence of this their distances could be overestimated.\\
On the other hand, \cite{Cornelisse} obtained an estimation of the range of $q$ and of M$_{NS}$, totally independent from assumptions about the distance to the source. This allows us to obtain a range of masses for the donor star of 0.46-0.50 M$_{\odot}$, assuming a mass of the neutron star of 1.4 M$_{\odot}$. \\
To sketch the most probable evolutive scenario, as well as to constrain the distance of XB 1254-690, we need to evaluate the companion star mass using the results of our timing analysis.
At this purpose, we assume that the companion star is a main sequence star in thermal equilibrium. The self-consistency of this hypothesis will be tested in the subsequent part of the discussion.
All the available X-ray data of XB 1254-690 clearly demonstrate that the source is a persistent X-ray emitter over a temporal window of about 26 years. As a consequence of this, we can properly assume that the companion star fills its Roche lobe, continuously transferring part of its mass to the neutron star. Thus, we impose that the companion star radius R$_{2}$ has to be equal to the Roche lobe radius R$_{L2}$, given by the expression of \cite{Pac}:
\begin{equation}\label{eq:Pacinzsky}
R_{L2}=0.46224\;a\left(\frac{m_{2}}{m_{1}+m_{2}}\right)^{1/3},
\end{equation}
where $m_{1}$ and $m_{2}$ are the NS and companion star masses in units of solar masses and $a$ is the orbital separation of the binary system. Hereafter, we are going to assume a NS mass of 1.4 M$_{\odot}$ for the subsequent analysis.
Our assumption that the companion star belongs to the lower main sequence brings us to adopt the mass-radius relation of \cite{Neece} for M-stars:
\begin{equation}\label{eq:m2_th}
\frac{R_{2}}{R_{\odot}}=0.877\;m^{0.807}_{2}.
\end{equation}
Using eq. \ref{eq:Pacinzsky} and eq. \ref{eq:m2_th} along with the third Kepler law, that links the orbital separation $a$ with the value of the orbital period found with the linear ephemeris, we obtain a mass of 0.42$\pm$0.04 M$_{\odot}$ for the donor star. Here we took into account an accuracy of 10\% in the mass estimation \citep[see][]{Neece}.\\
Our estimation of the donor star mass of 0.42$\pm$0.04 M$_{\odot}$ is in accordance with the values of \cite{Cornelisse} and as a consequence of this, we guess that the donor is probably a main sequence star in thermal equilibrium.
With this assumption, we can estimate the maximum distance that the system can have, assuming that the Kelvin-Helmholtz time-scale $\tau_{KH}$ (i.e. the characteristic time that a star spends to reach the thermal equilibrium) is equal to the mass transfer time-scale $\tau_{\dot{M}}$. 
 
The Kelvin-Helmholtz time-scale is given by the relation
\begin{equation}\label{eq:t_m}
\tau_{KH}=3.1\times10^{7}\left(\frac{M_{2}}{M_{\odot}}\right)^{2}\frac{R_{\odot}}{R_{2}}\frac{L_{\odot}}{L} {\rm yr}
\end{equation}
of \cite{Verbunt}, where we adopt the mass-luminosity relation for M-type stars of \cite{Neece}
\begin{equation}\label{eq:m_L}
\frac{L_{2}}{L_{\odot}}=0.231\;\left(\frac{M_{2}}{M_{\odot}}\right)^{2.61}
\end{equation}
and the mass-radius relation of \cite{Neece} of eq. \ref{eq:m2_th}.\\
On the other hand, the mass transfer time-scale is given by:
\begin{equation}\label{eq:tau_m}
\tau_{\dot{M}}=\frac{m_{2}}{\dot{m}}=\frac{G\;m_{1} m_{2}}{L_{X}\;R_{NS}},
\end{equation}
where $L_{X}$ is the bolometric source luminosity. \\
Imposing the similarity between $\tau_{KH}$ and $\tau_{\dot{M}}$ we obtain the maximum luminosity of the system in the hypothesis that the donor star is in thermal equilibrium in main sequence : $L_{X}\sim (10 \pm 2) \times 10^{36} \; {\rm erg \; s^{-1}}$.\\
The distance to the source for XB 1254-690 can be obtained by the flux inferred by \cite{Iaria_beppo} in the band 0.1-100 keV and from the X-ray luminosity just obtained as
\begin{equation}\label{eq:distance}
d=\sqrt{\frac{L_{X}}{4\pi \Phi}}= 7.6 \pm 0.8 \;{\rm kpc}.
\end{equation}
Once we obtained the new value of the distance, we repeated the comparison between the mass accretion rate predicted by the theory of secular evolution and that observed by \cite{Iaria_beppo}, rescaled for the updated distance.\\
In this case we observe that assuming $\beta$=1 (conservative mass transfer) there is no agreement between theory and observation adopting the parameter $f$=0.73 in eq. \ref{eq:mag_bra}. Using $f$=1.78, however, the agreement is achieved for $n\sim 0.8$ (see Fig. \ref{fig:confronto_ev}, left lower panel). In this case, therefore, we find a solution that is consistent with our initial hypothesis of a donor star in thermal equilibrium. Again the magnetic braking term T$_{MB}$ is necessary to explain the observed mass accretion rate, owing to the fact that otherwise for every $n$ no accordance is found between the theoretical mass accretion rate and the observed one.\\
For completeness, we also explore the non-conservative mass transfer scenario for both the distances considered above, adopting the same $f$ parameters used for the conservative case and assuming $n$=0.8. For a distance of 15.5$\pm$1.9 kpc the observation is in agreement with the theory for a value of $\beta \sim$ 0.9, i.e. the 90\% of the mass of the companion star is accreted onto the neutron star surface.
On the other hand, adopting a distance of 7.6$\pm$0.8 kpc we find a lower limit for $\beta$ of about 0.92. This actually means that most of the mass transferred by the donor star is accreted onto the neutron star.\\

In order to have a further confirmation of the scenario just depicted for XB 1254-690, and to understand the temporal evolution that the orbital separation of the system will undergo, we use the relation
\begin{equation}\label{eq:P_dot_Burderi}
\dot{m}_{-8}=87.5\;(3n-1)^{-1}m_{2}\left(\frac{\dot{P}_{-10}}{P_{2h}}\right),
\end{equation}
 of \cite{Burderi} to obtain a theoretical estimation of the orbital period derivative of the system.
In the equation, $\dot{P}_{-10}$ is the orbital period derivative in units of 10$^{-10}$ s/s, $P_{2h}$ is the orbital period of the system in units of two hours and $\dot{m}_{-8}$ is the secondary mass derivative in units of 10$^{-8}$ M$_{\odot}$ yr$^{-1}$.
Adopting the value of $\dot{m}$ at n=0.8, the mass of the companion star inferred in our analysis and the orbital period obtained with the linear ephemeris, we predict a negative $\dot{P}$ of -5$\times 10^{-13}$ s s$^{-1}$, meaning that the binary system is expected to shrink in agreement with the assumption of a companion star in thermal equilibrium and with the assumed mass-radius index of 0.8.\\
The theoretical value of $\dot{P}$ we inferred for this system is indeed compatible with the upper limit of the orbital period derivative obtained through the quadratic ephemeris: $|\dot{P}|<1.4\times 10^{-10}$ s s$^{-1}$.

\section{Conclusions}

In this work we update the existent orbital ephemerides for XB 1254-690, taking the advantage of about 26 years of X-ray data and performing direct measurements of the dip arrival times  on different pointed observations. We further increase the accuracy reached by \cite{Levine} by one order of magnitude.\\
The quadratic ephemeris, even though not statistically significant with respect to the linear ephemeris, allows to constrain the orbital period derivative of the system for the first time. Assuming that the companion star is in thermal equilibrium we infer a donor star mass of $0.42\pm0.04$ M$_{\odot}$, that is in agreement with the range of masses estimated by \cite{Cornelisse}, assuming that the mass of the neutron star is 1.4 M$_{\odot}$.\\
In our analysis we propose a different estimate of the distance to the source with respect to those obtained by \cite{Zand} and \cite{Galloway}. These authors provided an estimate of the distance to the source that, as they state, should be considered as upper limits to the distance to the source, owing to the relatively poor statistics of the data they analysed. We suggest a new distance of $7.6\pm0.8$ kpc that represents the distance for which the companion star has a Kelvin-Helmholtz time-scale that is similar to the mass transfer time-scale in this system. For larger distances, the inferred mass accretion rate would be higher, implying a mass transfer timescale shorter than the Kelvin-Helmholtz timescale of the donor, bringing the companion star out of thermal equilibrium. \\
These results, as well as the assumption of a neutron star of 1.4 M$_{\odot}$, allow us to state that the most probable scenario for this system is the one in which the companion star is in thermal equilibrium and that most of (if not all) the mass transferred by the companion is accreted onto the neutron star in a conservative way through the inner Lagrangian point, independently if we assume a distance of 15.5$\pm$1.9 kpc or of 7.6$\pm$0.8 kpc.
Moreover, the analysis strongly supports the idea that the magnetic braking plays an important role in the the angular momentum loss in this binary system. In the hypothesis that the companion star is in thermal equilibrium with a mass-radius index of about 0.8, we also predict that the binary orbit is shrinking at a rate of about -5$\times10^{-13}$ s s$^{-1}$. This prediction can be easily tested with future observations, when the uncertainty on the orbital period derivative we have now will be greatly reduced.

\section*{Acknowledgements}

This research has made use of data and/or software provided by the High Energy Astrophysics Science Archive Research Center (HEASARC), which is a service of the Astrophysics Science Division at NASA/GSFC and the High Energy Astrophysics Division of the Smithsonian Astrophysical Observatory.\\
This research has made use of the VizieR catalogue access tool, CDS, Strasbourg, France.\\
This work was partially supported by the Regione Autonoma della Sardegna through POR-FSE Sardegna 2007-2013, L.R. 7/2007, Progetti di Ricerca di Base e Orientata, Project N. CRP-60529.\\
We also acknowledge financial contribution from the agreement ASI-INAF I/037/12/0. AR acknowledges Sardinia Regional Government for the financial support (P.O.R. Sardegna F.S.E. Operational Programme of the Autonomous Region of Sardinia, European Social Fund 2007-2013 - Axis IV Human Resources, Objective l.3, Line of Activity l.3.1.).

\bibliographystyle{raa} 						
\bibliography{msgambino.bib}

\end{document}